\documentclass[review,3p]{elsarticle}
\usepackage[linesnumbered, ruled, noend]{algorithm2e}
\usepackage[hyphens]{url}
\usepackage{subfigure}
\usepackage{graphicx}
\usepackage{float}
\usepackage[normalem]{ulem}
\usepackage[colorlinks]{hyperref}

\AtBeginDocument{%
  \providecommand\BibTeX{{%
    \normalfont B\kern-0.5em{\scshape i\kern-0.25em b}\kern-0.8em\TeX}}}

\begin{document}


\title{Enabling Spatial Digital Twins: Technologies, Challenges, and Future Research Directions}


\author[inst1]{Mohammed Eunus Ali}

\affiliation[inst1]{organization={Department of Computer Science and Engineering, Bangladesh University of Engineering and Technology},
            addressline={ECE Building}, 
            city={Dhaka},
            postcode={1000}, 
            country={Bangladesh}}

\author[inst2]{Muhammad Aamir Cheema}
\author[inst1]{Tanzima Hashem}

\affiliation[inst2]{organization={Faculty of Information Technology, Monash University},
            addressline={
20 Exhibition Walk}, 
            city={Clayton},
            postcode={3164}, 
            state={VIC},
            country={Australia}}
            
\author[inst3]{Anwaar Ulhaq}
\affiliation[inst3]{organization={School of Computing, Charles Sturt University},
            city={Port Macquarie},
            postcode={2444}, 
            state={NSW},
            country={Australia}}

\author[inst4]{Muhammad Ali Babar}
\affiliation[inst4]{organization={School of Computer and Mathematical Sciences, The University of Adelaide},
            city={Adelaide},
            postcode={5005}, 
            state={SA},
            country={Australia}}

%

\begin{abstract}
  
A \emph{Digital Twin (DT)} is a virtual replica of a physical object or system, created to monitor, analyze, and optimize its behavior and characteristics. A \emph{Spatial Digital Twin (SDT)} is a specific type of digital twin that  emphasizes the geospatial aspects of the physical entity, incorporating precise location and dimensional attributes for a comprehensive understanding within its spatial environment. With the recent advancement in spatial technologies and breakthroughs in other computing technologies such as AI/ML, the SDTs market is expected to rise to $\$25$ billions covering a wide range of applications. The majority of existing research focuses on DTs and often fails to address the necessary spatial technologies essential for constructing SDTs. The current body of research on SDTs primarily concentrates on analyzing their potential impact and opportunities within various application domains.  As building an SDT is a complex process and requires a variety of spatial computing technologies, it is not straightforward for practitioners and researchers of this multi-disciplinary domain to  grasp the underlying details of enabling technologies of the SDT. In this paper, we are the first to systematically analyze different spatial technologies relevant to building an SDT in layered approach (starting from data acquisition to visualization). More specifically, we  present the key components of SDTs into four layers of technologies: (i) data acquisition; (ii) spatial database management \& big data analytics systems; (iii) GIS middleware software, maps \& APIs; and (iv) key functional components such as visualizing, querying, mining, simulation and prediction. Moreover, we discuss how modern technologies such as AI/ML, blockchains, and cloud computing can be effectively utilized in enabling and enhancing SDTs. Finally, we identify a number of research challenges and opportunities in SDTs. This work serves as an important resource for SDT researchers and practitioners as it explicitly distinguishes SDTs from traditional DTs, identifies unique applications, outlines the essential technological components of SDTs, and presents a vision for their future development along with the challenges that lie ahead.
\end{abstract}
\maketitle

\section{Introduction}
\label{section:introduction}

A digital twin is a virtual (or mirror) representation of a physical real-world entity, or system. According to Glaessegen et al.~\cite{glaessgen2012digital}, a digital twin is an integrated software system that mirrors the life of its corresponding physical object. With the rapid digital  technological growth in the last decade, the digital twins have received significant attention in many application domains including manufacturing, agriculture, healthcare, and smart \& sustainable cities. Major aims of these digital twins solutions are to improve the performance and efficiency of the system by real-time monitoring and predictive maintenance of physical entities, and by generating useful insights and giving feedback to the real-world entities for optimizing operational efficiencies. As the fastest growing sector of the fourth industrial revolution, the digital twins market is expected to grow from USD 12.7 billions in 2021 to USD 45 billions by 2026~\cite{WGIC2022}. Thus, we have witnessed a plethora of research and development involving digital twins in both academia and industries~\cite{kritzinger2018digital, tao2019make}.

The concept of digital twin was first introduced by NASA scientists to mirror the life cycle of space vehicle in 2012~\cite{glaessgen2012digital} and then later successfully applied in the field of manufacturing to improve the performance in manufacturing pipeline~\cite{kritzinger2018digital,shao2020framework}. After that, the digital twin technology  evolved in many directions -- one such area is smart and sustainable cities/precincts. Developing smart and sustainable cities is crucial as more than half of the world population lives in cities,  responsible for 70\% of the green house gas emissions\footnote{\url{https://www.worldbank.org/en/topic/urbandevelopment/brief/global-platform-for-sustainable-cities}}.  Thus, modeling urban environment and development, optimizing operational efficiencies, and enhancing decision-making are the keys to develop sustainable cities. To achieve the above goals, digital twins have evolved around geo-spatial context to give birth to a new type of digital twin, called \emph{spatial digital twins} in this paper. 

\subsection{Spatial Digital Twins (SDTs)} A spatial digital twin (SDT) essentially is a mirror representation of the real-world geospatial objects (e.g, buildings, roads) and systems (e.g., environmental or traffic monitoring). The geospatial consortium formally defines an SDT as a virtual representation of real-world entities and processes with precise location and dimensional attributes included in the model, where the virtual model is updated at a synchronized frequency~\cite{WGIC2022}.

At a high level, both SDTs and traditional digital twins (DTs) have many similarities, where both maintain virtual representation of real-world objects to assist monitoring, planning, and  decision making with predictive capabilities. The key differentiating factor of SDTs from DTs is that SDTs incorporate spatial context to capture and provide location and relative dimensional representation of geospatial objects and processes. 

More specifically, we envisage several key objectives of an SDT, which are: (i) to visualize, monitor, assess, and forecast the state and activities of different objects in a spatial region (e.g., monitoring and predicting energy consumption of buildings);  (ii) to predict the system response and to uncover previously unknown insights from the near real-time and historical data (e.g., prediction of flood in next 24 hours); (iii) to simulate what-if scenarios to uncover previously unknown insights  and to recommend corrective measure; and iv) to give feedback to the physical entities or processes to take necessary corrective measures (e.g., using real-time traffic feed, a digital twin system can control the timing of traffic lights to resolve/avoid traffic jams).

\begin{figure*}
\centering
\subfigure[Shadow effect]
{\includegraphics[width=0.36\textwidth]{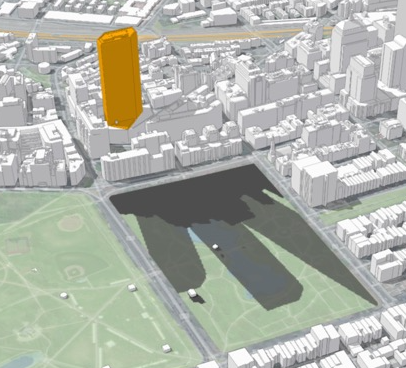}\label{fig:shadow}}
\hspace{3mm}
\subfigure[Fire rescue]
{\includegraphics[ width=0.38\textwidth]{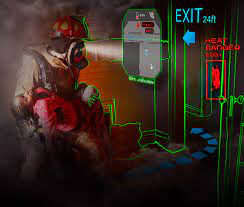}\label{fig:fire}}
\caption{Example Applications of Spatial Digital Twins}
\label{fig:mm_keeper}
\end{figure*}

\subsection{Applications}
While SDTs are often mistakenly confused with 3D modeling and visualization techniques used for urban planning and development, the capabilities of SDTs extend far beyond these applications, encompassing a broad range of functionalities with numerous potential uses. We have listed some important application areas as follows. 

\begin{itemize}
    \item \textbf{Urban Planing and Management:} 
    The spatial digital twin of a city can be used for monitoring and managing urban planning and development processes. For example, Boston city digital twin\footnote{\url{https://www.esri.com/about/newsroom/blog/3d-gis-boston-digital-twin/}} is used to evaluate how much shadow a proposed high-rise building will cast on a given park area (see Figure~\ref{fig:shadow}).

    \item \textbf{Natural Disaster:} In recent years, the scale and frequency of natural disasters such as floods and bushfires have seen a significant increase worldwide, largely attributed to the effects of climate change. For example, in March 2020, bushfire in Australia killed  3 billion animals, burnt more than 18 million hectors of land, and destroyed over 3000 houses\footnote{\url{https://en.wikipedia.org/wiki/2019-20_Australian_bushfire_season}}. Using spatial digital twins of forests, animals, buildings, and other infrastructure, a better rescue and fire containment plan can be made, which can  significantly reduce the extent of lives and property damages. An example of SDT-based fire rescue system\footnote{\url{https://thegoodnewstvshow.com/how-ar-firefighting-masks-improve-situational-awareness/}}
 is shown in Figure~\ref{fig:fire}.

   \item \textbf{Virtual and Augmented Reality Based Service:} SDTs hold tremendous potential in the realm of virtual reality, enabling users to simulate and interact with virtual models incorporating real-world entities. Additionally, they offer exciting possibilities in augmented reality, allowing users to seamlessly interact with and manipulate physical elements of the real world. For example, an electrician can see the layout of the wiring of the building retrieved from the SDT before repairing the fault. 
    
\item \textbf{Greenhouse Gas Emissions and Climate Change:}
Greenhouse Gas Emissions (GHG) is one of the major factors causing climate change and has a profound impact on our environment. To reduce GHG, which comes from fossil fuel consumption and other indirect source like electricity generation, it is important to monitor, manage, and plan for both macro and micro level energy consumption. SDTs can be helpful to monitor and manage current energy consumption, predict future consumption, and optimize the use of renewable energy sources in buildings, suburbs, or entire cities.
By leveraging SDTs, we can work towards reducing GHG emissions, promoting sustainability, and fostering the transition to a more environmentally conscious energy landscape.

\item \textbf{Intelligent Transportation: } SDTs can be used to operate on-demand vehicles on optimized routes, improve road safety, and reduce congestion by real-time monitoring and predictive analysis.

\item \textbf{Epidemiology and Public Health:} We have witnessed an unprecedented loss of lives in COVID-19. Also, we have seen that this disease can be better managed by applying advanced data analytics and AI techniques on data collected from various sources~\cite{rizvi2021clustering}. An SDT can be an assistive tool for spatio-temporal prediction of the infectious disease, identifying the source of the diseases, tracking the movement of infected patients, and making a better plan for containing the disease.

\item \textbf{Smart or Precision Agriculture:} Smart or precision agriculture that relies on sensing technologies to get real-time agricultural field data and AI\&ML technique for decision making can be immensely benefited by the SDT developed around crop cultivation, distribution, and market demand.
\end{itemize}

\subsection{Different Components of SDTs}

As SDTs need to integrate a vast array of spatial and geolocation data to accurately and comprehensively represent a large physical environment, the scale and complexity of an SDT is much higher than that of a traditional DT. For example, SDTs consist of various form of data that include location data, 3D models, spatial networks, satellite imagery, etc., and to acquire these data we need the support of positioning devices, Internet of Things sensors, drones, satellites, etc. The choice of data sources and categories will be determined by the specific use case and application of the SDTs. ~\cite{semeraro2023digital}.
Thus, to build an SDT, we need various spatial technologies. In this paper, we have identified four major building blocks of an SDT and discuss relevant spatial technologies for each of these building blocks (see Figure~\ref{fig:sdt} and Section~\ref{section:problem_definition} for details): 
   \begin{enumerate}
 \item  Data Acquisition and Processing: The first stage in developing an SDT is acquiring and processing data about the physical environment. This can include satellite imagery, LiDAR scans, aerial photography, IoT sensor data, and GIS data, among other forms of spatial and geolocation data. The data must be processed to create a digital representation of the physical environment.  

 \item  Data Modeling, Management, and Processing: This the second layer that models, stores, and provide data management and processing platform for spatial data. This include modeling different data types such as location, 3D, image, etc., and provide data management and processing platform such as PostgreSQL, geoSpark, etc.

 \item GIS Software, Maps, and APIs: This middle layer contains GIS platforms and base maps, and available APIs to perform different operations on spatial data. This layer fetches the data from the previous layer and fusion it with maps,  GIS, or 3D platform software. 

 \item Key Functional Components: This final layer involves key functionalities that include simulation, predictive modeling, visualization, query processing, etc. To provide these functionalities, an SDT need the support the spatial technologies of previous layers, and  other supportive technologies like AI \&ML, block-chain, etc.
\end{enumerate}

\subsection{Scope of This Work and Contributions}
Most of the existing works related to SDTs highlight different application areas and use cases of live city digital twins (e.g.,~\cite{white2021digital, ClemenALOOSG21, lei2023challenges, tao2019make, dembski2020urban}). A recent work~\cite{mylonas2021digital} lists a number of city digital twins and discusses commercial software tools for building 3D city digital twins, which include ArcGIS by ESRI~\cite{arcgisDT}, Azure Digital Twins platform by Microsoft~\cite{azureDT}, 3DEXPERIENCE platform by Dassault Systemes~\cite{3dexpDT}, etc. 
Our paper is orthogonal to these software systems as we focus on key spatial technologies that can be used to build an SDT, and these software systems can also be used as a base with other spatial technologies for a quick development of SDTs.
In particular, in this paper, we primarily focus on summarizing and analyzing various spatial technologies including spatial data management system, big data analytic system, GIS software, APIs \& Tools, etc., to develop an SDT, and highlight the potential future works in this domain. The key contributions of this paper are as follows.
\begin{itemize}
    \item We highlight the potential differences between SDTs and traditional DTs, and also identify key application areas of SDTs.
    \item We are the first to propose a layered framework to identify and categorize existing spatial technologies and map  them to different layers of building blocks of SDTs.
    \item We analyze the roles of emerging technologies such as AI/ML, blockchain, and cloud computing in the context of SDTs
    \item Finally, we identify a number challenges and potential future research directions to guide  researchers and practitioners working on SDTs.  
\end{itemize}

The rest of the paper is organized as follows. Section~\ref{section:related} presents related work and case studies for some real-world spatial digital twins. Building blocks of SDTs are discussed in Section~\ref{section:problem_definition}. In Section~\ref{sec:relevant_tech}, role of other relevant technologies such as AI/ML, blockchain and cloud computing in SDTs is discussed. Section~\ref{sec:futurework} discusses some key challenges and opportunities in this area. The paper is concluded in Section~\ref{sec:conclusion}.

\section{Related Work}
\label{section:related}


The concept of digital twin emerged in the early 2000s~\cite{glaessgen2012digital} and has since evolved to include various forms such as spatial digital twin~\cite{WGIC2022}, mobility digital twin~\cite{WangGHWGAT22,FanYYJC0S22}, urban digital twin, human digital twin, and others. In Section~\ref{sec: DTvariants}, we discuss different variants of digital twins and in In Section~\ref{sec: case_studies}, we discuss several case studies for SDTs. 


\subsection{Digital Twins and Variants}
\label{sec: DTvariants}
In~\cite{VanDerHornM21}, the generalized characteristics of a digital twin are identified in terms of three primary components: a physical reality, a virtual representation, and interconnections that exchange information between the physical reality and virtual representation. A recent report~\cite{WGIC2022} focuses on the importance of integrating spatial data with digital twin, and describes a spatial digital twin as 
a holistic dimensional and location-based representation of assets, infrastructure and systems. Mobility digital twin and and smart city digital twin are examples of spatial digital twin. Other types of digital twin include industry digital twin, health digital twin and others.

\textbf{\textit{Mobility digital twins.}} In~\cite{WangGHWGAT22}, the authors developed a holistic framework for mobility digital twins that manage mobility entities like human, vehicle and traffic. In~\cite{FanYYJC0S22}, the authors proposed a fine grained human mobility prediction system for mobility digital twins. Mobility digital twins are one example of spatial digital twins as one of their main aims is to visualize and predict the movement locations of mobility entities in the virtual space.   

\textbf{\textit{Smart city digital twins.}} Digital twin has been widely explored in the context of smart cities. Smart city digital twin is also known as urban digital twin. Key challenges for implementing digital twin for smart cities come from the complexity of systems and variant end user types~\cite{MylonasKKAAM21}.~\cite{ClemenALOOSG21} shows that city planners, policy stakeholders, and other decision-makers can be highly benefited from digital twin for smart cities, which is simply formed by integrating real-time data with an existing multi-agent framework. ~\cite{xia2022study} presents the detailed literature study on the integration of BIM and GIS data in the context of city digital twin. They also provide theoretical analysis and design the framework, and highlight  future research directions in GIS and BIM integration field. ~\cite{dollner2020geospatial} highlights the potentials of machine learning techniques for 3D point cloud processing for building geospatial digital twins. ~\cite{shahat2021city} presents a review of existing works on city digital twins and also identifies the current and prospective potentials and challenges of digital twin cities.


\textbf{\textit{Industry digital twins.}} Digital twins have been considered for manufacturing automation in industries before they are applied for smart cities. Researchers have focused on adopting digital twin technologies from various dimensions for engineering product family design and optimization~\cite{LIM202082} and production optimization~\cite{MIN2019502}. A recent study~\cite{ASSADNETO2021178} identifies and assesses the importance of key features (i.e., the digital modeling feature, the analytics support feature, the timeliness update feature, and the control feature) of digital twin for the success of its implementation manufacturing. .

\textbf{\textit{Health digital twin.}}
A human digital twin~\cite{10063703}, also known as health digital twin virtually models the life cycle of a human with an aim of smart health management. In~\cite{SHENGLI2021100014}, the authors proposed to use augmented reality assisted digital twin to model human digital twin. Health digital twins have improved personalized health care~\cite{VenkateshRK22} due to the availability large scale data for the disease risk and progression prediction process.


\textbf{\textit{Others.}} The concept of digital twin has been also implemented for indoor space~\cite{AhujaSPXOA021} to capture gaze behavior of a class. In~\cite{WANG2022100014}, the authors studied the integration of digital twin technologies in four levels of energy sector applications (i.e.,low carbon city, smart grid, electrified transportation and advanced energy storage system), identified their challenges and discussed possible solutions.

\subsection{Spatial Digital Twin Case Studies}
\label{sec: case_studies}
In this section, we present some live city digital twin systems from different part of the world, which will help us to understand the scope and relevant technologies of SDTs. 
\subsubsection{Singapore}
Singapore has created a 3D digital platform, called Virtual Singapore\footnote{\url{https://www.nrf.gov.sg/programmes/virtual-singapore}}, that provides a semantic 3D model of the city. The model includes detailed information such as texture, material representation of geometrical objects, terrain attributes, and models of buildings that encode their geometry and components down to their fine details. The platform integrates data from various public agencies and co-ordinates it with existing geospatial and non-geospatial platforms to enrich the 3D model with 2D data and other information. Advanced information and modelling technology allows this spatial digital twin of Singapore to be infused with different sources of static, dynamic, and real-time city data and information, such as demographics, movement, and climate.

Singapore digital twin offers four major capabilities: virtual experimentation; virtual test-bedding; planning and decision-making; and research and development. The twin can be used for virtual test-bedding or experimentation, such as examining the coverage areas of 3G/4G networks and providing realistic visualisation of poor coverage areas. It can also be used as a test-bedding platform to validate the provision of services, such as modelling and simulating crowd dispersion in the 3D model of the new sport hub to establish evacuation procedures during an emergency. With its rich data environment, Singapore digital twin is a holistic and integrated platform to develop analytical applications for planning and decision-making, such as analysing transport flows and pedestrian movement patterns. The platform's rich data environment also allows researchers to innovate and develop new technologies or capabilities, such as advanced 3D tools.

The digital twin has several potential use cases in tackling liveability issues. For example, the platform can be used for collaboration and decision-making by integrating various data sources, including data from government agencies, 3D models, information from the Internet, and real-time dynamic data from IoT devices. The digital twin allows different agencies to share and review the plans and designs of various projects in the same vicinity. Another potential use case is enhancing emergency response capabilities by modelling and simulating crowd dispersion and establishing evacuation procedures during an emergency. The digital twin can also be used to improve transportation planning by analysing transport flows and pedestrian movement patterns. 

\subsubsection{Zurich, Switzerland}

The Digital Twin project for the city of Zurich~\cite{schrotter2020digital} is a cutting-edge initiative aimed at creating a virtual representation of the city's physical environment. The project uses advanced technologies such as sensors, drones, and 3D modeling software to collect and integrate data on buildings, infrastructure, and public spaces. This data is then transformed into a 3D spatial model that creates a digital twin of the city. The digital twin serves as a virtual replica of the city, allowing planners, architects, and engineers to test and visualize different urban design scenarios, predict the impact of proposed changes, and improve the efficiency of urban management. The key features of the digital twin include the ability to simulate urban climate scenarios and analyze the interactions between the built environment and various urban systems. The project enables detailed visualization of urban planning scenarios, as well as concrete civil engineering and construction projects, including detailed building projects using BIM models. The platform also enhances collaboration between internal and external stakeholders, while leveraging technologies like virtual and augmented reality to provide more vivid visualizations of proposed projects. Ultimately, the digital twin project for Zurich is expected to enhance the city's sustainability, resilience, and livability while fostering innovation and collaboration

\subsubsection{NSW, Australia}

The New South Wales (NSW) spatial digital twin\footnote{\url{https://nsw.digitaltwin.terria.io/}} is a 3D virtual model of the New South Wales state in Australia. It is an innovative initiative of the NSW Government that uses cutting-edge technology to create a digital representation of the state's physical and geographical features.
 The digital twin is based on a range of data sources, including satellite imagery, LiDAR data, and other geographical data. These data sources are combined to create a high-fidelity, 3D model of the state's infrastructure, built environment, natural environment, and other assets. The model is constantly updated in real-time as new data becomes available.

The NSW spatial digital twin is designed to support a range of applications and use cases. For example, it can be used to visualize and analyze urban planning scenarios. The 3D model can be used to visualize different urban planning scenarios, such as the impact of new buildings, roads, and other infrastructure projects. This can help planners and decision-makers to make informed decisions about development projects and their impact on the environment and community. It can also be used to monitor and manage infrastructure assets such as roads, bridges, and utilities. It can provide real-time information on the condition of these assets and help identify maintenance needs and potential risks. This digital twin can also be used to support emergency management and response efforts. For example, it can be used to simulate and analyze the impact of natural disasters, such as bushfires and floods, and help emergency responders to plan and coordinate their response efforts.

The NSW Government is also exploring other potential use cases for the digital twin, such as supporting environmental monitoring and management, tourism, and cultural heritage preservation. Overall, the NSW Spatial Digital Twin is a powerful tool that has the potential to transform the way the state's resources and assets are managed and utilized.

\subsubsection{Boston, USA}

Boston has created a comprehensive digital twin of the city that incorporates a wide range of data, including building layouts, public transit routes, tree canopies, and proposed and under-construction buildings. The city began working on the twin in 2005, using the 3D modeling capabilities of geographic information system (GIS) software and the expertise of Esri professional services.
The twin allows for testing of planning decisions before implementation in the real world, helping planners visualize the world in its existing form. It also includes tools for analyzing shadows and evaluating the impact of new zoning and development. This helps the Boston Planning \& Development Agency (BPDA) ensure that projects adhere to the city's zoning code height, density, and usage requirements.
The qualitative analysis of the twin provides a quick visual to see shadows cast by any new building, while quantitative assessments provide more in-depth measurements, such as the extent and duration of shadows through the seasons. The city's open data, including parcel ownership, zoning districts, historic landmarks, and open space, is integrated into the 3D model to analyze and evaluate city planning and development. BPDA uses planning and shadow tools to create real-world visualizations for a wide variety of decision-making tasks, including planning and development, flood modeling, shadow studies, and line-of-sight evaluation.

\section{Building Blocks of Spatial Digital Twins}
\label{section:problem_definition}

\begin{figure*}[hbtp]
    \centering
\includegraphics[scale=0.6]{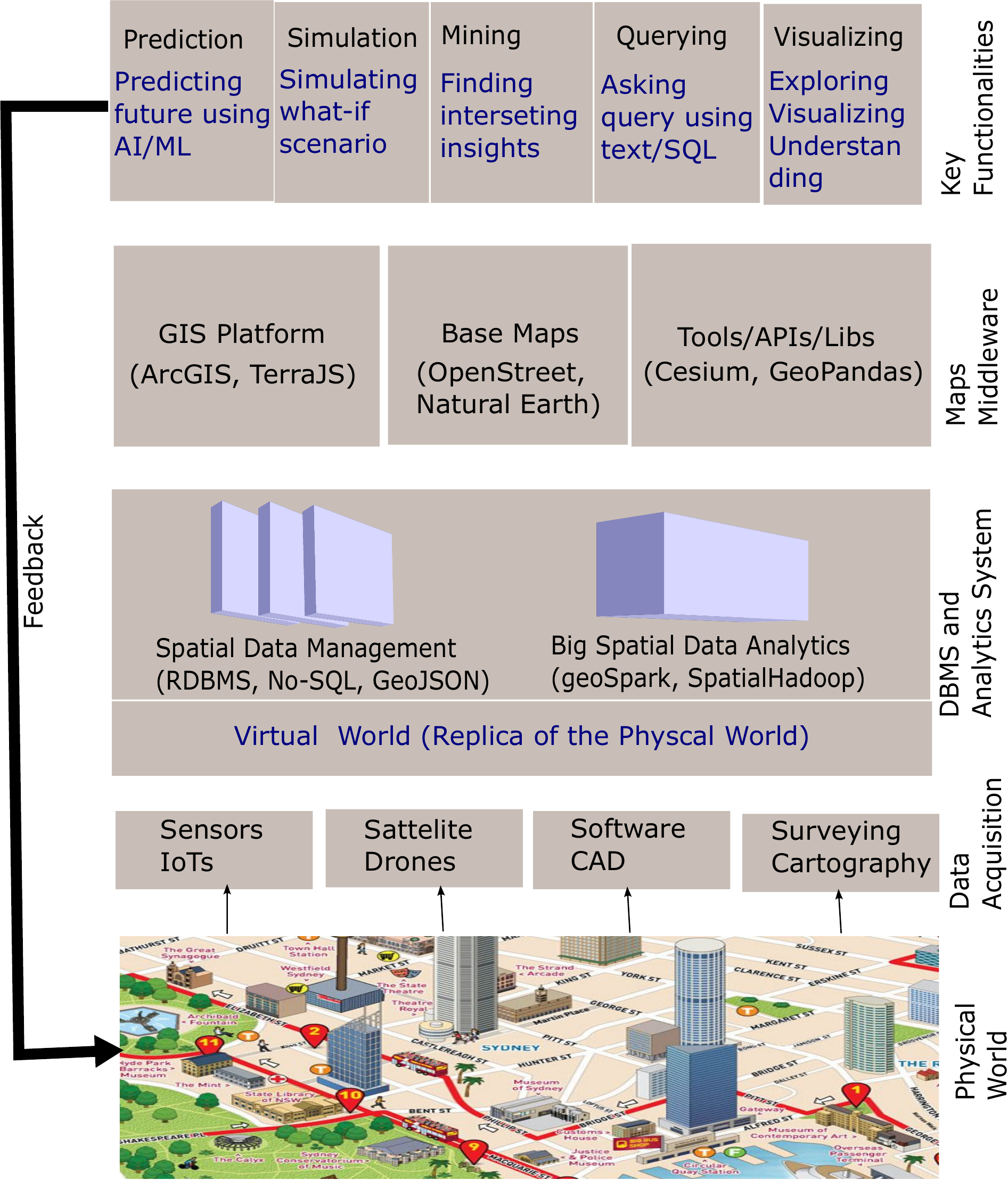}
    \caption{A block diagram of different key components of a spatial digital twin (SDT).}
    \label{fig:sdt}
\end{figure*}

In this section, we discuss enabling technologies for building a spatial digital twin (SDT). Figure~\ref{fig:sdt} shows the block diagram of different layers of technologies and tools required for building an SDT. These layers are: i) data acquisition and processing technologies that include sensors, satellites, software systems etc. (Section~\ref{subsec:dataacq}); ii) spatial data modeling, storage and management systems  (Section~\ref{subsec:dm}) and big spatial data analytics systems (Section~\ref{subsec:bda});  iii) GIS and map-based middlewares (Section~\ref{subsec:maps}); and iv) Key functionalities and/or operations (Section~\ref{subsec:func}). It is important to note that there is no strict vertical or horizontal ordering of the spatial technologies as presented in Figure~\ref{fig:sdt}, i.e., vertical components of a layer can interact with each other, and similarly, a top layer component can interact with a bottom layer component skipping the middle layers.

\subsection{Data Acquisition and Processing}
\label{subsec:dataacq}

The first stage in developing an SDT is acquiring and processing data about the physical environment. This includes  remote sensing technologies (e.g., LiDAR, drones, satellite imagery etc.), positioning technologies (e.g., GPS), Internet of Things (IoT) sensors (e.g.,  air quality sensor, traffic sensor, GPS, cameras, etc.), software systems (e.g., CAD). The data captured from various data acquisition technologies must be processed to create a digital representation of the physical environment.





 Building SDTs involves capturing a variety of real-world geo-spatial objects of different spatio-temporal dimensions. Examples include 2D or 3D layouts of buildings, spatio-temporal traces of  moving objects, satellite images of the landscape, spatial networks (e.g., road networks), time series data of energy consumption of a building etc. The acquisition process of these data also varies significantly ranging from IoT-enabled sensor devices for capturing evolving data such as changing locations or energy consumption  to software system such as CAD for generating detailed 3D layouts of buildings. As the acquisition technologies widely vary across data types and the required level of representation necessary for the spatial digital twins, it is challenging to make a uniform standard and quality of the captured data~\cite{li2022spatial}. Consider an example of a GPS device, which is used to track and update the locations of a moving object. The frequency of updates of a GPS device decides the quality of the data that can be used in different applications, e.g., a low frequency update may be enough to derive the traffic conditions of the road, but may not be sufficient to analyze how eco-friendly the route is as the calculation of an eco-route depends on many detailed factors~\cite{fahmin2022improving} such as acceleration/deceleration, instantaneous velocity, slope of the road, etc.
Similarly, 3D modeling of objects can also be captured in many different resolutions, e.g., drones are used to capture 3D point clouds~\cite{jo2021dense} which then pass through different pre-processing modeling steps~\cite{to2021drone} (i.e., segmentation, approximation, etc.) to generate 3D layouts of different buildings of the city. However, if we need the detailed indoor layout of a building we need to resort to software systems like CAD.  

Existing research~\cite{botin2021digital} highlights different technologies and different layers of spatial data such as terrain, buildings, infrastructure, and mobility for creating an urban digital twin. Recent research also identifies several data acquisition challenges including the large varieties of data, interoperability, and the requirements of huge computing power~\cite{dembski2020urban}.



\subsection{Data Modeling, Storage and Management}
\label{subsec:dm}

The  spatial data captured from data acquisition layer needs to be modelled and then stored in a data management system. Broadly, in geospatial domain, we can categorize spatial data modeling into two groups: raster data models, and vector data models~\cite{samet2006foundations}. 
In raster data models, the whole space is divided into grid-cells and each cell is represented as a pixel denoting the cell information. A matrix data structure (or an array of grid cells) is generally used for storing the raster data. This is the most common data model of the GIS community. Continuous geospatial data such as satellite image, thematic map, etc. are generally modeled using raster data model. 
In contrast, in vector data models, spatial entities are represented using spatial primitive geometric data types such as points, lines, and polygons. For example, the location of a school is represented as a point, a suburb can be represented as a ploygon, and a road network can be represented as a graph. Each object can be associated with other non-spatial attributes, e.g., a school has many non-spatial attributes like, name, ranking, number of students, category, etc. This vector modeling is the most commonly used paradigm in the spatial database community.

A key component of an SDT is the data management system as it needs to handle a wide range of data with varying levels of spatial and temporal granularity. Over the years, many spatial data management techniques and algorithms have been introduced to handle various spatial data types such as points, lines, plygons and time series~\cite{chen2008open}.

To handle large spatial data types and operations, traditional  relational database management systems (RDBMS) have extended their scopes to handle spatial data. For example, Oracle Spatial~\cite{oracle2000oracle}, Microsoft SQL Server~\cite{aitchison2009beginning} and PostgreSQL/PosGIS~\cite{strobl2008postgis} have provided support for many important spatial data types and some key spatial operations. PogreSQL has also included the support of spatial raster data (e.g., images)~\cite{guliato2009postgresql}. As reported in recent studies~\cite{shukla2016comparing, sveen2019efficient}, PostgreSQL has the best spatial support in terms of spatial data types, queries, and scalability. It supports different geometry types that include 2D and 3D geometries such as points and ploylines, and supports various queries such as joins and k-nearest neighbors (kNNs). Other popular spatial RDBMSs (e.g., Oracle Spatial and SQL Server) are also continuously adapting new features to make these systems suitable for the era of big spatial data. Recent research~\cite{yao20183dcitydb} also discusses how 3D city models can be managed and stored using these spatial RDBMSs.

Unstructured data such as text, documents and graphs are augmenting spatial data in many applications.  NoSQL (Not-Only-SQL) database~\cite{davoudian2018survey} systems have become very popular to handle large volumes of unstructured data of various types due to their schema-free and 
scalable natures. Many such NoSQL database systems have recently been extended to handle spatial data types and operations~\cite{agarwal2016performance, baas2012nosql, ben2016spatial}.

For example, MongoDB, a database for managing documents, provide support for basic GeoJSON objects, (e.g., points, linestring, and polygons). Similarly, Oracle NoSQL, a key-value store database, also supports common geometry objects,
and a set of spatial operators (e.g., intersect, inside, near, etc.) for processing spatial data~\cite{agarwal2016performance}. Neo4j, a popular graph database management system, has a spatial extension called Neo4j Spatial~\cite{baas2012nosql} that can store, index, and
process spatial data.

In summary, though both RDBMS and NoSQL database systems are continuously adopting spatial features, still there is no comprehensive evaluation of how these systems perform for different spatial operations on complex spatial objects such as 3D buildings. As for the SDTs, after building the underlying database, there will be significantly more read operations than write operations, and we may not need to strictly follow ACID properties of the transactional database system, and a combination of RDBMS and NoSQL can be the best options for handing different use-case scenarios.

Apart from RDBMS and NoSQL, there are some popular file formats such as GeoJSON, Shapefiles, GPX and Keyhole Markup Language (KML) for storing and sharing geospatial data in different formats~\cite{gisformats}. Among them GeoJSON is the most common format for storing and sharing spatial objects in the forms of points, lines, ploygons, etc. Similarly, the shapefile format was created by Esri, the company that develops ArcGIS software, for storing and sharing spatial objects. The GPX format is an open standard XML markup language for storing a collection of timestamps and latitude/longitude coordinates of moving objects. The KML format is commonly used to store two- and three-dimensional geographic data.


\subsection{Big Data Analytics System}
\label{subsec:bda}

The rapid and extensive use of mobile location-based apps, GPS-enabled cars, autonomous vehicles, UAVs, IoT devices, satellite imagery, and more has led to an unprecedented generation of spatio-temporal data. For example, each day, approximately a billion tweets are created, with 30-40\% of them containing geolocation information. An autonomous vehicle at the lower end of the autonomous spectrum produces about 1.4 terabytes data per hour\footnote{\url{https://blogs.sw.siemens.com/polarion/the-data-deluge-what-do-we-do-with-the-data-generated-by-avs/}}. To host this large volume of spatio-temporal data in an SDT, and supporting various analytical and query operations, recent research focuses on the spatial extensions of big data analytics systems such as Hadoop~\cite{shvachko2010hadoop}, Spark~\cite{salloum2016big}, and NoSQL~\cite{davoudian2018survey}. These big data systems can make the SDT system scalable through distributed processing. 

Hadoop~\cite{shvachko2010hadoop} uses MapReduce framework for distributed processing of big data. As there is no support for spatio-temporal data in Hadoop, a number of extensions such as Hadoop-GIS~\cite{aji2013hadoop} and SpatialHadoop~\cite{eldawy2015spatialhadoop} have been developed to handle spatial data types such points, trajectories, etc. As Hadoop is a disk-based system, the performance can deteriorate with large number of I/O operations. To solve this problem, main-memory based big data systems such as SpatialSpark~\cite{yu2019spatial}, GeoSpark~\cite{hagedorn2017big} etc. have been developed. Another group of big spatial data processing systems has been emerged using NoSQL paradigm, e.g., MD-HBase~\cite{nishimura2011md}, TrajMesa~\cite{li2020trajmesa}, etc.

As spatio-temporal data is growing at an unprecedented rate and the SDTs need to host different data types to find insights and to help making predictive decisions, we need the help of these big spatial data processing systems in building an SDT. However, unlike the database management systems that have decades of research behind them and have been extensively tested, these big spatial data processing systems are still at their growing phase and there is no benchmark study on the performance of these systems using various spatial data types and data streams. 


\subsection{Maps and GIS Based Middleware}
\label{subsec:maps}

In this section, we discuss software and tools that are being used for visualizing spatial (or geographical data) on a map, and performing various spatial operations with ease. We have categorized these tools and software into three groups: GIS software; basic maps; and APIs \& libraries.
These systems are essential for smart, quick  and efficient development of an SDT. Next, we present the details.

\subsubsection{GIS Software}
An integrated geographical information system (GIS) provides an environment that integrates maps to visualize and interact with spatial data, and tools to perform various operations on those data. They also provide platforms to find interesting patterns and insights from the data and visualize on a map. Some widely used GIS software are ArcGIS (from ESRI)~\cite{kennedy2013introducing},  and QGIS (open source)~\cite{kurt2016mastering}. These GIS software have the support for all the basic spatial data operations and analytics facilities and visualize on the map. These can also be integrated with DBMSs and other underlying spatial data analytics platforms. Also such software has programming support for different languages like python, R, etc. Boston digital twin uses ArcGIS technology to build their SDTs~\cite{boston3d}.

As one of the major feature of an SDT is to visualize and interact with the urban environment, an opensource 3D globe platform, Cesium~\cite{cesium} has been developed. Cesium enables developers to build geospatial applications around 3D maps, Cesium is particularly optimized for visualizing large datasets, such as terrain, buildings, and real-time sensor data, and for creating immersive geospatial experiences. CesiumJS is an open source JavaScript library for creating world-class 3D globes and maps with the best possible performance, precision, visual quality, and ease of use. New South Wales digital twins~\cite{nswtwin} use Cesium for building their spatial digital twin.

\subsubsection{Map services}

At a broad level, we can view existing map services like OpenStreetMap, Bing Maps, or Google Maps as spatial digital twins. These platforms create and utilize digital representations of physical road networks, rivers, points of interest, and more to address various queries, such as locating a point of interest or finding a route to a destination. A number of different layers of maps including satellite maps, traffic maps, terrain maps, 3D maps are also widely available in these map services. In addition to these basic layers that can be used from different map services as an API, an SDT needs to have a number of other map layers such as utility map layer, indoor map layer, or any other utility-based layer such as greenhouse gas emissions or energy usage. Thus map layers play a vital role in developing an SDT. Hence, one can use these widely used map services as a base to build their own SDT.

\subsubsection{Tools, APIs and Libraries}
So far, we have seen a wide range of spatial technologies such as RDBMSs, GeoJSON files, GIS software, and spatial big data analytics platforms that are used to manage and process spatial data. An SDT relies on the integrated capabilities of many of these platforms. Thus, to build an SDT, we need a strong support of libraries and APIs of different programming languages to integrate the SDT platform with the standard GIS and spatial technology related platforms and perform various processing, analyzing, mining, and visualizing operations. A variety of APIs and libraries with spatial capabilities are available in almost every popular language like C++/Java, Python, etc. 

Low-level libraries such as JTS~\cite{jts}, GEOS~\cite{geos}, Spatial4j~\cite{spatial4j} have been developed using C++ or Java  for modeling and processing spatial data (e.g., creating vector geometry). These foundational libraries for spatial data modeling are being used to build GIS software, big spatial data processing and analytic systems, etc. On the other hand, with the popularity of high-level languages, like Python and R, for big data analytics and processing, a number of APIs and libraries are available. These libraries are primarily designed for various spatial data related tasks including I/O optimized data fetching, processing spatial data using different spatial algorithms, performing analysis using statistical and advanced ML, visualizing different types of spatial data. We briefly discuss some of the Python libraries and APIs because Python seems to be the most popular high-level language for big data processing in the era of AI \& ML.

GeoPanda~\cite{geopanda} is a spatial extension of Pandas, a popular Python library for data science. 
Scipy~.~Spatial~\cite{scipyspatial} has libraries for spatial algorithms and data structures such as nearest neighbor search algorithm. Python Spatial Analysis Library (PySAL)~\cite{PySal} is designed for spatial data analysis tasks such as clustering, hot-spot analysis, correlation, visualisation, etc. It includes support to handle both vector and raster spatial data. To manage and process trajectory data, Python has traja~\cite{traja} library. Apart from the above libraries, there are many other spatial libraries that can be useful for developing an SDT (e.g., see~\cite{westra2015python} for details).

Apart from the above software and technological tools, some commercial software solutions~\cite{mylonas2021digital} have been recently developed by combining many of the above spatial tools \&techniques, which can also be  used as a starting base to develop an SDT. Microsoft Azure Digital Twins platform~\cite{azureDT} enables one to create a digital twin in the form of knowledge graph, which allows developers to create a digital twin data model to represent real assets and their relationships with each other. They also allow seamless integration with IoT devices for continuous data feed. The 3DEXPERIENCE~\cite{3dexpDT} platform (by Dassault Systemes) provides tools for creating 3D models and also a number of simulation tool for simulating real life objects and their processes. The HxDR platform~\cite{hxdr} (by Hexagonis) is an SaaS platform that focuses mainly on providing the necessary tools creating 3D replicas of urban environments. OpenCities Planner~\cite{opencities} by Bentely Systems is a city-scale SDT planning and visualizing application software.

\subsection{Key Functional Components}
\label{subsec:func}
We have broadly divided the functionalities of an SDT into five different categories: exploration \& visualization; querying; mining; simulation; and predictive analysis. Next, we discuss these in details.

\subsubsection{Exploration and Visualization}
One area where SDTs have evolved the most is visualization. Many of the research papers and systems~\cite{nswtwin} focus on visualization, and also refer the visualization system as the main goal of the spatial digital twins (which is not entirely a valid claim as the other key functionalities discussed in this section are equally important). In the visualization of SDTs, usually a basic map layer is used and different layers of data are added such as 3D buildings, mobility traces and spatial networks. Besides maps and GIS software like OpenStreetMap and ArcGIS, there are a number of 3D visualization platforms such as Cesium~\cite{cesium}, 3DEXPERIENCE~\cite{3dexpDT}, HxDR platform~\cite{hxdr}, etc. While certain aspects of spatial insights and relationships can be grasped, the current visualization technologies fail to capture the full extent of profound spatial correlations within diverse types and timelines of data. For instance, visualizing a 3D building along with its time series energy consumption on a map proves to be a challenging task.


\subsubsection{Querying}
Despite extensive research on spatial databases and spatial queries, only a handful of commercial database systems, like Oracle Spatial and PostgreSQL, currently offer support for spatial queries on the underlying spatial database. However, these systems do not cater to emerging data types such as 3D data and trajectories, nor do they address specialized queries like identifying visible 3D objects from a query point~\cite{arman2017vizq} or finding similar trajectories~\cite{ali12maximum}. Most importantly,
it is not easy for the users to write spatial SQL queries and retrieve desired information from these database systems. 
Users commonly find it more convenient to formulate spatial queries in the form of text, such as searching for a point of interest (POI) on mapping platforms like Google Maps and Bing Maps. This integration of textual queries with underlying spatial technologies, specifically in an SDT, can prove to be a highly effective method for users to access the spatial data they desire. Recent NLP research focuses on converting user textual description to SQL (aka, text-to-sql) and retrieve desired information for relational databases~\cite{kumar2022deep}. Spatial extensions of such techniques can help users querying spatial data. 

\subsubsection{Mining}

Data mining techniques are critical in  analysing the data generated by SDTs and identifying patterns, relationships, and trends that can provide valuable insights into how the physical world is functioning. For example,  mining techniques can be used to identify areas in a city where greenhouse gas emissions are high or to detect anomalies in the air quality of certain areas. It is important to design techniques that can identify interesting insights~\cite{ding2019quickinsights,martinez2020data,demiralp2017foresight} from the data automatically with little to no human intervention. Such techniques are essential to the effective use of SDTs because these allow us to turn raw data into actionable insights that can be used to optimise the performance of physical systems and improve decision-making. Later, in Section~\ref{sec:challenges:insights}, we discuss some research challenges in data mining for SDTs that must be tackled to fully exploit their potential.


\subsubsection{Simulation}

Simulations are crucial in a spatial digital twin as they provide a non-destructive method of testing and analysing the performance and behaviour of a system. Advancements in technology, such as cloud computing, machine learning, and Internet of Things (IoT), have revolutionised simulation functionalities in SDTs~\cite{boschert2016digital}.

Spatial digital twin simulation operations can be applied in numerous scenarios. For example, traffic flow can be simulated in transportation, and transportation routes can be optimised to reduce congestion and improve safety~\cite{biswas2023safest}. Energy consumption can be simulated in a building or neighbourhood, and energy-saving opportunities can be identified. Production processes can be simulated to enhance efficiency and minimise waste. Emergencies, such as natural disasters or terrorist attacks, can be simulated to plan and train for response efforts. In healthcare, patient flows, and hospital operations can be simulated to optimise resource allocation and improve patient care \cite{ agalianos2020discrete, schluse2016simulation}.

Various software tools are available for simulations of SDTs, including AnyLogic, OpenStudio, Simio, and Arena Simulation Software. These tools can be used to model and simulate systems in different applications, such as transportation, manufacturing, healthcare, and logistics~\cite{umair2021impact}. Real-time information is a critical component of simulation operations in SDTs, as it provides current and accurate data about the system being modelled. Real-time information can be acquired from sensors, cameras, and other IoT devices, improving the accuracy and relevance of the simulations \cite{ schluse2018experimentable}.

\subsubsection{Prediction}

AI/ML can utilise real-time data to predict and provide feedback for SDTs~\cite{gao2022guest}. This is achieved by training models on historical and real-time data to forecast future outcomes, identify anomalies, and provide recommendations for improvement. Real-time data is critical in several scenarios, such as traffic flow optimisation, predictive maintenance, and emergency response planning \cite{hasse2019digital, ruppert2020integration,he2023intent}. For example, in traffic flow optimisation, real-time data from sensors and cameras can be used to predict traffic congestion and optimise transportation routes to reduce travel time and improve safety~\cite{alam2023serverless}. In predictive maintenance, real-time data from sensors and devices can detect equipment anomalies and predict maintenance needs, avoiding costly downtime.

Several organisations have implemented AI/ML in SDTs with real-time data. For instance, San Diego uses real-time data to predict and prevent wildfires by analysing weather conditions and other environmental factors.  Dubai Electricity and Water Authority uses AI/ML to predict energy consumption and optimise energy usage in buildings, leveraging real-time data to respond to fluctuations in energy demand. The Port of Rotterdam uses real-time data to predict ship movements and optimise port resources, enabling more efficient and effective port operations~\cite{jagannath2022digital, baranda2021aiml}.

\section{Other Relevant Modern Technologies}
\label{sec:relevant_tech}

The recent advancements in AI and ML technology, along with breakthroughs in blockchain and cloud computing have significantly enhanced our ability to efficiently solve a wide range of problems across various domains. Next, we discuss the details of how these technologies affect the development and operation of SDTs.

\subsection{AI \& ML}
AI/ML technologies have a significant role to play in the development and operation of DTs and SDTs~\cite{lv2022artificial}. SDTs are complex systems that require advanced technologies to optimise their performance and efficiency. AI/ML technologies are essential components in this regard as they enable digital twins to learn from data, predict future outcomes, and provide recommendations for improvement \cite{gao2022guest, baranda2021aiml}. There are three main aspects of this contribution. 
\begin{itemize}

\item AI/ML technologies can optimise various processes in spatial digital twins \cite{ hasse2019digital}. For example, in smart buildings, AI/ML can predict energy consumption and optimise energy usage to reduce costs and emissions. In smart cities, AI/ML can optimise transportation routes, reduce congestion, and enhance road safety~\cite{sohail2023data}. In manufacturing, AI/ML can optimise production processes, reduce waste, and increase efficiency.

\item AI/ML technologies can also be used for predictive maintenance in spatial digital twins    \cite{hosamo2022digital}. By analysing real-time data from sensors and other devices, AI/ML can predict when maintenance is required for equipment, thus avoiding costly downtime and improving the longevity of assets.

\item In addition to optimising processes and maintenance, AI/ML can also enhance the user experience of spatial digital twins via enhanced recommendations \cite{onile2021uses}. AI/ML can personalise user experiences, predict user preferences, and improve overall user satisfaction.
\end{itemize}

\subsection{Blockchain}

SDTs, being data-driven software applications, are anticipated to witness an increased utilization of Distributed Ledger Technologies (DLTs), commonly referred to as Blockchain, to enhance the trustworthiness of data within them.
Blockchain is a distributed system that is considered as either a distributed ledger or a replicated state machine \cite{TranBB21}.
DLTs offer fundamental capabilities that can greatly benefit SDTs, such as trustworthy data acquisition, processing, and storage.  A blockchain enabled system supports a network of computers to store, validate and update an append-only transaction log (i.e., a ledger). Ethereum, Hyper Fabric and Bitcoins are some of the implementations of DLTs that can be considered for their suitability for providing trustworthiness of SDTs. However, there are several other points of concerns to be evaluated before incorporating DLTs in SDTs. 

While blockchains’ functionality and attributes can provide a reliable support system to ensure trustworthiness, it is important to assess whether or not and which blockchains are suitable for SDTs. Some areas of exploration can be the suitability of the types of blockchain, e.g., private/public, the nature of consensus protocols, e.g., Proof-of-Work (PoW) or Proof-of-Authority (PoA), and empirically understanding the characteristics of blockchain networks that store and operate blockchains. For example, performance, resource consumption, and availability are the key attributes to be studied as these are not fixed characteristics of blockchains networks but can vary based on how a blockchain network is designed for which network design knowledge becomes quite vital. There is an increasing amount of scientific and commercial literature and readily available products that can assist in any kind of feasibility study and/or experimentation for systematically assessing the mechanisms and potential benefits of using DLTs for SDTs.

\subsection{Cloud Computing}
SDTs are expected to be supported by large scale technological infrastructure for computing, storage and networking. The support infrastructure is also expected to fulfil the quality attributes requirements such as scalability, latency, security and availability.With the maturity and widespread adoption of virtualized infrastructure technologies such as Cloud/Fog, containers, and Network Function Virtualization (NFV)~\cite{Prokhorenko2020}, SDTs can effectively harness the potential of cloud computing-based technologies. Considering the plethora of cloud services offered by public cloud providers like Amazon Web Services and MS Azure, as well as private cloud infrastructures, it becomes crucial to comprehensively comprehend the various technological and business models available for acquiring and utilizing virtualized infrastructure services to support SDTs. For example, technological models include Infrastructure-as-a-Service (IaaS), Platform-as-a-Service (PaaS) and Software-as-a- Service (SaaS). 
Cloud technologies primarily offer three main deployment models for data storage, processing, and networking: private, public, and hybrid. All the technological and business models have their own pros and cons. For example, public cloud can be quite expensive as compared to private cloud, which is also considered relatively more secure. However, an organisation opting private cloud is expected to have a variety of knowledge and expertise for designing, implementing and operating private cloud that meet the functional and non-functional requirements of SDTs; the use of commercial cloud infrastructure does not entail such requirements of knowledge and expertise. 


\section{Challenges and Future Work}
\label{sec:futurework}
Considering the current state of knowledge in spatial technologies and SDTs, we have identified a key set of challenges and opportunities that needs immediate attention from researchers and practitioners in order to build a sustainable SDT. In this section, we discuss these challenges and list some important directions for future work.

\subsection{Multi-modal and Multi-resolution Data Acquisition}
Most of the existing research~\cite{lei2023challenges} in this domain highlights the data acquisition and integration as one of the major challenges in SDT. An SDT involves acquisition of a wide variety of spatial and associated non-spatial data. As an SDT needs to use a wide range of devices to capture data of different spatial and temporal resolution, scale, and  precision, quality of these data largely varies. To the best of our knowledge, no longitudinal study has been done for bench-marking data capturing devices used to capture various spatial data so that the data integration from difference sources/devices can be done seamlessly. For example, the integration of BIM and 3D GIS data still remains a challenge due to the generation process (or data sources) and the differences in the standards used in these two formats.


\subsection{NLP for Spatial Queries}
Current query processing techniques on SDTs is limited to running SQL queries on relational or NoSQL database systems or running textual queries on map services (e.g., finding a POI on Google Maps). Recent breakthroughs in NLP enable researchers to devise text-to-SQL techniques that facilitate automatic translation of natural language text to SQL and run the query to retrieve answers from database tables\footnote{\url{https://paperswithcode.com/task/text-to-sql}}. Although the accuracy of such approaches is still not good enough for commercial use, the recent breakthroughs in very large conversational generative language models, such as GPT-3/4~\cite{gpt4} and LaMDA~\cite{lambda}, are showing great promise in natural language based query processing in database systems. We envision that there is a huge scope of research in  natural language based query processing in SDTs by exploiting the power of these very large language models. As the spatial data and their relationships, and other associated data describing spatial entities make the whole data interaction use cases complex, it would be interesting to see how we can use different spatial (e.g., adjacent/nearby) and structural (e.g., road network) properties along with tabular data to answer user queries on SDTs.

\subsection{Benchmarking the Databases and Big Data Platform for SDT}
Some recent research works experimentally evaluate different spatial databases and big data platforms but under limited settings. In~\cite{shukla2016comparing}, the authors compared Oracle Spatial and PostgreSQL performance using a small spatial dataset consisting of  New York city census blocks and streets, where they used select and range query to measure the performance. In~\cite{makris2021mongodb}, the authors compared MongoDB and PostgreSQL to measure the performance for spatio-temporal range and proximity queries, where they used polygons and vessel movement (a sequence of lat-long pairs) datasets. Another recent work~\cite{shin2022comparative} compared Geospark based system against three database technologies, namely MongoDB, PostgreSQL, and Amazon EC2 services where they also used polygons and vessel movement datasets to measure the performance of range and proximity queries. As we have observed from the current research,  database technologies in this domain are still not mature and only support  basic spatial data and spatial queries. As an SDT generally hosts various forms of data ranging from 3D building data to continuous stream of energy consumption, effectiveness of handing these data in existing platforms has not been benchmarked yet.

\subsection{Automated Spatial Insights}\label{sec:challenges:insights}

Spatial digital twins generate vast amounts of data, often sourced from a wide variety of origins. It is critical to be able to automatically identify interesting insights from such data without the need of a human input. Furthermore, the techniques that can predict future behavior, risks, opportunities and trends are also important so that appropriate actions can be taken. While automatically identifying insights has been studied~\cite{ding2019quickinsights,martinez2020data,demiralp2017foresight}, none of these techniques are designed specifically for the spatial data. Therefore, these techniques cannot provide spatial insights which are crucial for the operation and management of spatial digital twins. Spatial insights can take various forms, such as detecting neighborhoods with abnormally high greenhouse gas emissions, uncovering spatial correlations between different attributes such as air quality and traffic accidents in various parts of a city, or mapping and highlighting regions with above-average crime rates and their correlation with seemingly disparate attributes such as electricity usage, waste production, and traffic patterns.
Furthermore, temporal aspects of spatial data must also be considered in generating insights. For example, it might be interesting to study how the spatial correlation between two or more attributes evolves over time. Unfortunately, the existing techniques cannot be applied or easily extended for spatial digital twins due to their inability to consider spatial features.  Furthermore, it is crucial to design techniques that are efficient so that the insights can be generated in a timely manner, allowing system operators to intervene promptly if necessary

\subsection{Multi-modal Analysis}
The integration of image and text in multimodal language models, such as CLIP~\cite{clip}, enables them to learn the data space jointly and effectively address challenges related to multimodal data. Furthermore, the remarkable advancements in multimodal large generative models, like GPT-4~\cite{gpt4}, have demonstrated significant potential in the field of multimodal learning. We envision that it is possible to answer many queries and to find interesting insights from the captured satellite image/drone image (i.e., raster spatial data) with other spatial and non-spatial data. More specifically, combining different forms of data such as real-time data (e.g., traffic, energy consumption) collected by sensors, images taken by a satellite or a drone, spatial neighborhood features, and train a large multimodal model may be able to generate useful insights.

\subsection{Building Simulation Environment}
As many factors such as social interaction, economic factors, and human factors may not be possible to be captured in the SDTs, it is important to build a simulated environment, where these factors can be simulated so that their association with the captured SDT data can be assessed. Future research should focus on how to build a realistic simulated environment tailored for the SDTs. Different simulation software like AnyLogic, OpenStudio, and Simio, have been developed for applications such as transportation, logistics, and manufacturing. As the scope and scale of an SDT are significantly different from that of DTs, there is a potential avenue for research on how to develop a platform for simulating different factors involving an SDT.

\subsection{Visualizing Complex and Diverse Interactions}

An SDT involves various forms of data such as infrastructure data (e.g., 3D buildings, roads, etc.), sensor data (e.g., energy/gas/water consumption, traffic, etc.), and social media data (e.g., Twitter, Instagram, etc.). There are  a number of data visualization challenges involving the interaction of these complex data objects. For example, as the visualization of more than three dimension is not comprehensible for typical users, it is hard to combine different forms of data, e.g., 3D building with time series energy consumption of the building on a map. Also as the data can be of higher dimension and of different types, visualizing correlations among those datasets across spatial and temporal dimension needs more research. Also, how to add and visualize different layers of data attached to a particular location on a map or GIS software to observe interesting insights and interaction of the associated data needs to be investigated.

\subsection{Mitigating the Security and Privacy Concerns}
Researchers have emphasised on the importance of integrating security  and privacy solutions in digital twins to ensure the quality of services like providing data-driven decisions~\cite{Charitonidou2022,9362182}. Reliability, trust, transparency, integrity, authenticity, anonymity and selective disclosure are some of the security and privacy aspects that need attention to make DTs a success. The security and privacy challenges that have been identified for DTs in general also apply for SDTs. Besides, the use of location data imposes additional security and privacy concerns for the SDTs as location data may act as an identifier of an individual and may reveal sensitive information~\cite{HashemK07,MokbelCA06}. For example, although mobility data of an individual allows an SDT to monitor traffic in indoor and outdoor spaces or contact tracing~\cite{liu2023contact}, continuous sharing of locations of an individual will allow others to know the individual's movement trajectory and infer the places visited by the individual. If a place represents an individual's office then the individual would be identified, and if a place is a liver clinic then it may reveal the individual's sensitive health information. Researchers have developed techniques like obfuscation, $k$-anonymity and space transformation that trad-off between utility and privacy of location data. 

There are a few works~\cite{LvQLYW22,PervezKGS23,RAMU2022} that proposed countermeasures to prevent security and privacy attacks for DTs. In these works, blockchain technology~\cite{LvQLYW22}, federated learning models~\cite{RAMU2022}, and cryptographic protocols~\cite{PervezKGS23} have been shown as techniques to address various security and privacy concerns. Considering that location data may raise new threats, future research should focus on identifying new security and privacy attacks, designing solutions to protect them, and investigating the applicability of existing security and privacy solutions for SDTs. 




\section{Conclusion}
\label{sec:conclusion}
In this paper, a thorough and organized collection of spatial technologies is presented, forming the fundamental components of Spatial Digital Twins (SDTs). The building blocks of SDTs are categorized into four layers, and a comprehensive overview of key spatial technologies is provided for each layer, covering aspects such as data acquisition, processing, storage, and visualization. We have also presented how modern technologies like AI \&ML, blcokchain, and cloud computing can facilitate more efficient and effective SDTs. It is important to note that there is currently no existing study that specifically focuses on identifying these crucial technologies essential for the development of SDTs. Consequently, researchers and practitioners working in this multidisciplinary domain, particularly those with limited knowledge of geo-spatial advancements, may encounter difficulties in adopting these technologies for SDTs. Therefore, this interdisciplinary work holds immense potential in bridging the gaps among researchers and practitioners in fields such as geo-spatial, urban and transport engineering, and city planning.




\bibliographystyle{elsarticle-num} 
\bibliography{main}



\end{document}